\documentclass{ws-procs9x6}

\begin{document}

\title{STABILITY CONDITIONS IN GAPLESS SUPERCONDUCTORS}

\author{E. GUBANKOVA}

\address{Center for Theoretical Physics, Massachusetts Institute 
of Technology,\\ Cambridge, MA 02139, USA\\
E-mail: elena1@mit.edu}

\begin{abstract}

Gapless superconductivity can arise when pairing occurs between fermion species with
different Fermi surface sizes, provided there is a sufficiently large mismatch between
Fermi surfaces and/or at sufficiently large coupling constant. In gapless states, secondary
Fermi surfaces appear where quasiparticle excitation energy vanishes. This work focuses on homogeneous and isotropic superfluids in the $s$-wave channel, with
either zero (conventional superconductor), one, or two spherical Fermi surfaces. The stability conditions for these candidate phases are analyzed. It is found that gapless states with one Fermi surface are stable in the BEC region, while gapless states with two Fermi surfaces are unstable in all parameter space. The results can be applied to ultracold fermionic atom systems.
\end{abstract}


\bodymatter

\section{Introduction}

This work focuses on stability conditions and their possible relations to different Fermi surface topologies in a superconductor with unequal number densities of fermions (or unequal chemical potentials)\cite{1}. The physics of paired fermion systems with unequal densities of the two fermion species is of interest in the study of two physical systems:  (1) Ultracold atomic fermionic gases, where one can freely choose populations in two hyperfine states of the fermionic atom. Experimental work is currently being conducted with these systems. (2) Quark matter in the interior of neutron stars, which is believed to be a color superconductor. There, the mismatch in quark Fermi surfaces is driven by differences in quark masses and electric and color neutrality conditions. In both cases, the ground state of such a system is the subject of current debate. The study of fermionic superfluids with imbalanced populations is a novel subject, because until recently, experimental evidence for this situation was elusive. In a superconductor, an imbalance in population between spin-up and spin-down electrons can be created by a magnetic field. However, due to the Meissner effect, the magnetic field is either completely shielded from the superconductor bulk or enters in the form of quantized flux lines or vortices.

Cold fermionic atomic gases provide new possibilities in the experimental exploration of fermionic superfluids with unequal mixtures of fermions. In cold atoms, Feshbach resonance permits tuning of the interaction from (1) attraction and a resulting superfluid of loosely bound pairs (the BCS regime at $\zeta>n^{-1/3}$); to (2) repulsion and a resulting Bose-Einstein condensate of tightly bound molecules (the BEC regime at $\zeta<n^{-1/3}$), where $\zeta$ is the size of a pair and $n^{-1/3}$ is the interparticle separation (or, more precisely, the mean free path). With equal mixtures, the BEC--BCS crossover is smooth, with no phase transition. With asymmetric densities, one or more phase transitions are expected, and a more complex phase diagram may result.

In the weak coupling regime, a BCS superfluid remains stable as long as the difference in chemical potential is small compared to the pairing gap, $\delta\mu<\Delta$; the gap prevents the excess unpaired atoms from entering the superfluid state. By either increasing the mismatch or reducing the binding energy (and hence decreasing the gap), a quantum phase transition from the superfluid to normal state takes place, and superflidity ceases. The point at which the phase transition occurs is known as the Clogston limit, which can be estimated as $\delta\mu\sim\Delta\sim\mu\exp(-1/g)\ll\mu$, where $\mu$ is the Fermi energy. Thus only an exponentially small population imbalance is allowed in weak coupling. In strong coupling, as one approaches Feshbach resonance, the situation is quite different. On the BEC side, superfluidity with imbalanced population is robust over a wide range of parameter space. Surprisingly, on the BCS side, macroscopic imbalance is also possible in a superfluid state, possibly due to the formation of a gapless superfluid at strong coupling. This gapless superfluid incorporates large numbers of unpaired fermions which reside at the secondary Fermi surfaces. Here, we consider stability conditions for gapless states with different Fermi surface topologies.

\section{Definitions for the screening masses and susceptibility.}

We consider two species of nonrelativistic fermions, $\psi=(\psi_1,\psi_2)$ with the same mass but with different chemical potentials, $\mu={\rm diag}(\mu_1,\mu_2)$. There is an attractive interaction only between different species of fermions, $g\left(\psi^{\dagger}\sigma_2 \psi^*\right)\left(\psi^T\sigma_2\psi\right)$. The order parameter is $\Phi^{\dagger}=\Delta\sigma_2$ where $\Delta=2g<\psi^T\sigma_2\psi>$, it defines pairing in the singlet channel, spin $0$, $[2]\times [2]=[1]+[3]$. The order parameter breaks the original group associated with conservation of particle numbers of each species down to the diagonal subgroup associated with conservation of difference in particle numbers, $U(1)_{\alpha_1}\times U(1)_{\alpha_2}\rightarrow U(1)_{\alpha_1-\alpha_2}$ which is invariant under simultaneous rotation $\alpha_1=-\alpha_2$. Total number of particles is not conserved. Formally, one can express this pattern of symmetry breaking by gauging the theory. We introduce two external gauge fields, and couple each species of fermions to its gauge field, $\psi_1$ couples to $A_1$ with $g_1$ and $\psi_2$ couples to $A_2$ with $g_2$, that is reflected by the generators of the gauge group, $T_1={\rm diag}(1,0)$ and $T_2={\rm diag}(0,1)$. According to the symmetry breaking pattern, gauge fields mix in a superconductor, rotated sum of fields $\tilde A_1=A_1+A_2$ is screened with non-zero Meissner mass via Andersen-Higgs mechanism, $m_M^2\neq 0$ corresponding to $U(1)_{\alpha_1+\alpha_2}$, and rotated difference of fields $\tilde A_2=A_1-A_2$ propagates in a superconductor unscreened, Meissner mass is zero $m_M^2=0$ corresponding to $U(1)_{\alpha_1-\alpha_2}$. There is a different mixing of gauge fields in case of Debye masses. In QCD, with color and electromagnetic gauge groups, $SU(3)_c\times U(1)_{EM}$, mixing of gauge fields is more complicated.

We treat four-fermion interaction on a mean field level. We integrate over the
fermions, obtain fermion determinant,
\begin{equation} 
{\cal Z} = \int {\cal D}A\,\exp\left[S_A + 
\frac{|\Delta|^2}{4g}-\frac{1}{2}{\rm Tr}\ln({\cal S}^{-1} + {\cal A})\right]\,.
\label{part}
\end{equation}
We use the Nambu-Gorkov formalism with particle-hole basis, $\Psi=(\psi,\psi^*)$. Inverse fermion propagator and gauge field are $4\times 4$ matrices in this basis, including the Nambu-Gorkov and two fermion species indices,
\begin{equation} 
{\cal S}^{-1} \equiv \left(\begin{array}{cc} [G_0^+]^{-1} & \Phi^- \\ \Phi^+ &
[G_0^-]^{-1}
\end{array}\right) \, ,
\end{equation}
where the inverse free fermion propagators are
$[G_0^\pm]^{-1} = i\partial_t \pm \frac{\nabla^2}{2m} \pm \mu$,
and we have abbreviated for the gauge fields 
${\cal A}={\rm diag}(A^+, A^-)$ with
$A^\pm = \pm \Gamma_a A_a^0 \mp\frac{\Gamma_a^2}{2m}\,{\bf A}_a^2
 -\, \frac{i\Gamma_a}{2m}(\nabla\cdot{\bf A}_a+
{\bf A}_a\cdot\nabla)$
and $\Gamma_a=g_a T_a$.
Performing derivative expansion and collecting terms quadratic in the gauge
field, we produce fermion loops, $\Pi_{ab}^{00}$, $\Pi_{ab}^{i0}$,
$\Pi_{ab}^{0i}$, $\Pi_{ab}^{ij}$. Debye mass in one-loop is defined by the
temporal component of polarization operator, Meissner mass is given by the
spatial component,
$m_{D,ab}^2\equiv -\lim_{{\bf p}\to 0}\Pi_{ab}^{00}(0,{\bf p})$ and
$m_{M,ab}^2\equiv \frac{1}{2}\lim_{{\bf p}\to
0}(\delta_{ij}-\hat{p}_i\hat{p}_j)\Pi_{ab}^{ij}(0,{\bf p})$,
where $\hat{p}_i\equiv p_i/p$. Screening masses are given\cite{1} 
\begin{eqnarray} 
m_{D,ab}^2&=& -\lim_{P\to 0}\frac{1}{2}\frac{T}{V}\sum_K {\rm Tr}[S(K)\Gamma_a^-S(K-P)\Gamma_b^-] \, , 
\nonumber\\
m_{M,ab}^2&=& \frac{1}{2m}\lim_{P\to 0}\frac{T}{V}\sum_K \left(\phantom{\frac{k^2}{m}}\hspace{-0.5cm}
\delta_{ab}{\rm Tr}[S(K)\bar{\Gamma}_a^2]\right. 
\label{screening} \\ 
&+&\left.\frac{k^2}{2m}\,[1-(\hat{p}\cdot\hat{k})^2]\,{\rm Tr}[S(K)\Gamma_a^+S(K-P)\Gamma_b^+]\right)\, ,
\nonumber
\end{eqnarray}
where $S$ is the fermion propagator and we have introduced the
following matrices in Nambu-Gorkov space,
$\Gamma_a^{\pm}= {\rm diag}(\Gamma_a,\pm \Gamma_a)$ and 
$\bar{\Gamma}_a^2= {\rm diag}(\Gamma_a^2, - \Gamma_a^2)$.
Physically, Debye mass to all loops, including 
ladder diagrams, is equivalent to compressibility of a system, and Meissner mass
can be associated with the density of superconducting fermions. 
These $2\times 2$ matrices in two-fermion space shall be evaluated in the following section 
in order to obtain stability conditions for gapless superconductors. 

One may derive the pressure from the partition function in 
(\ref{part}) using the Cornwall-Jackiw-Tomboulis formalism. 
The pressure is the negative of the effective potential at its stationary
point (i.e., with the propagators determined to extremize the effective
potential). The
fermionic part of the pressure is
$p = \frac{1}{2}\frac{T}{V}{\rm Tr}\ln{\cal S}^{-1} + \frac{1}{2}\frac{T}{V}{\rm
Tr}[{\cal S}_0^{-1}{\cal S} -1]
+\Gamma_2[{\cal S}]$,  
where ${\cal S}_0 = {\rm diag}(G_0^+,G_0^-)$ is the tree-level fermion
propagator in Nambu-Gorkov
space and $\Gamma_2[{\cal S}]$  is
the sum of all
two-particle irreducible diagrams.
The number densities is defined $n_a = \partial p/\partial \mu_a$.
The number susceptibility $\chi$ is defined as the derivative of the number density with respect to the chemical potential (at constant volume and temperature). Using the expression for the pressure, we obtain\cite{1}
\begin{eqnarray} \label{susc}
\chi_{ab}=\frac{\partial n_a}{\partial \mu_b}&=&  -\frac{1}{2g_ag_b}\frac{T}{V}\sum_K{\rm Tr}[\Gamma_a^-{\cal S}(K)\Gamma_b^-{\cal S}(K)]
\nonumber\\
&-&\,\frac{1}{2g_a}\frac{T}{V}\sum_K{\rm Tr}\left[\Gamma_a^-{\cal S}(K)\frac{\partial \Sigma(K)}
{\partial\mu_b}{\cal S}(K)\right] \, ,
\end{eqnarray}
where $\Sigma$ is the fermion self-energy, ${\cal S}^{-1} = {\cal S}_0^{-1} + \Sigma$.
The first term on the right-hand side of this equation
is given by the one-loop result for the electric screening mass, cf.\ Eq.\ (\ref{screening}).
For the second term, we assume that the self-energy $\Sigma$ depends on $\mu$ only through the 
gap, then we obtain\cite{1}
\begin{equation} \label{chidef}
\chi_{ab} = \frac{m_{D,ab}^2}{g_ag_b} +
\frac{\partial n_a}{\partial \Delta}\frac{\partial \Delta}{\partial \mu_b} \, .
\end{equation}
In general, the self-energy $\Sigma$
contains terms of any number of fermion loops. Consequently, the number susceptibility
contains terms of arbitrary many fermion loops too, corresponding to
the exact Debye mass including all possible perturbative insertions.
Remarkably, the free fermion result for $\chi$, i.e. $\Sigma=0$, gives the
one-loop result for $m_D^2$. Susceptibility is equivalent to compressibility
of a system.
We shall use Eq.\ (\ref{chidef}) in the following section to compute the number susceptibility.
As this equation shows, it goes beyond the one-loop result for the electric screening mass.

\section{Results}

\begin{figure*} [ht]
\begin{center}
\hbox{\includegraphics[width=0.33\textwidth]{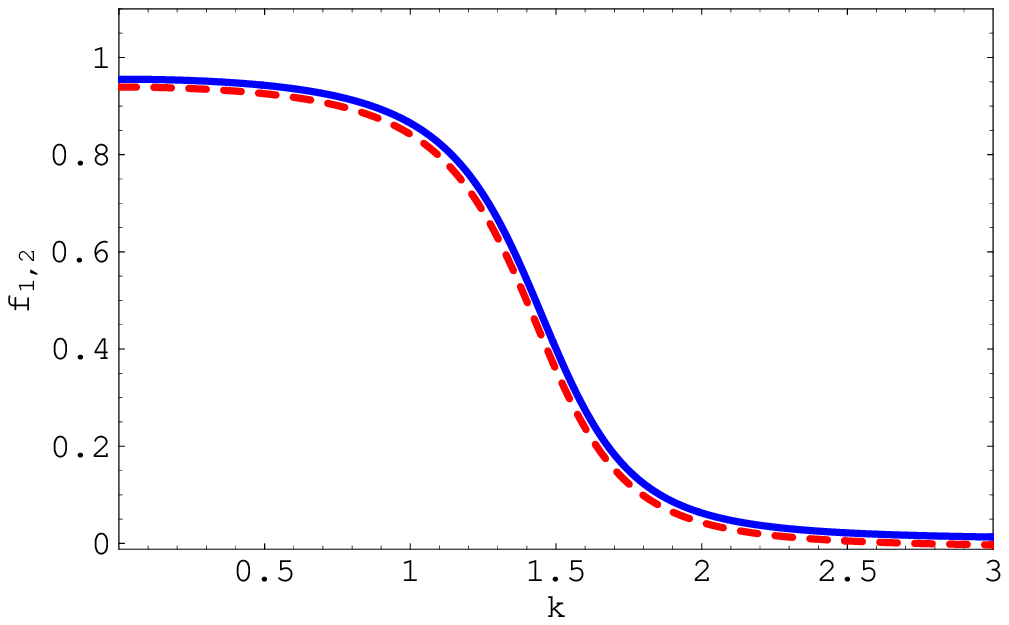}
\includegraphics[width=0.33\textwidth]{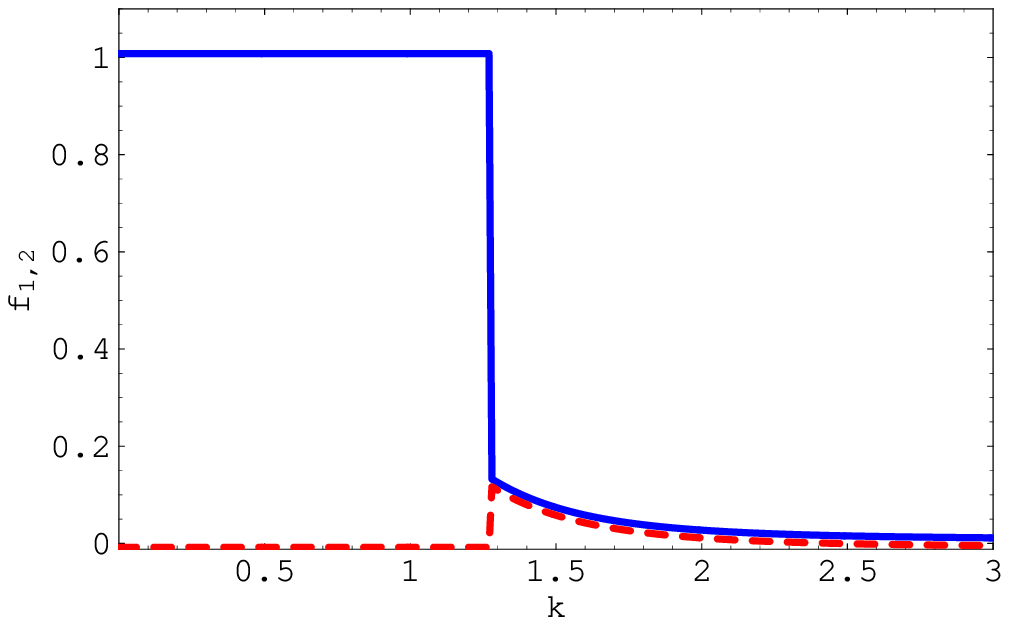}
\includegraphics[width=0.33\textwidth]{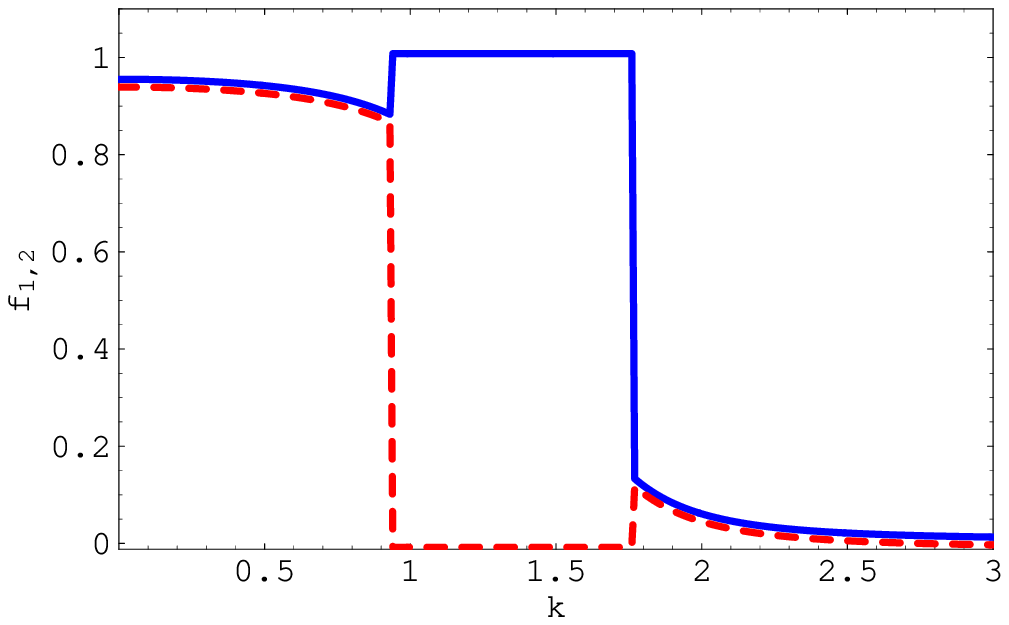}}
\caption{Schematic plot of possible quasiparticle occupation
numbers (in arbitrary units).}
\label{figoccupation}
\end{center}
\end{figure*}

\begin{figure} [ht]
\begin{center}
\hbox{\includegraphics[width=0.5\textwidth]{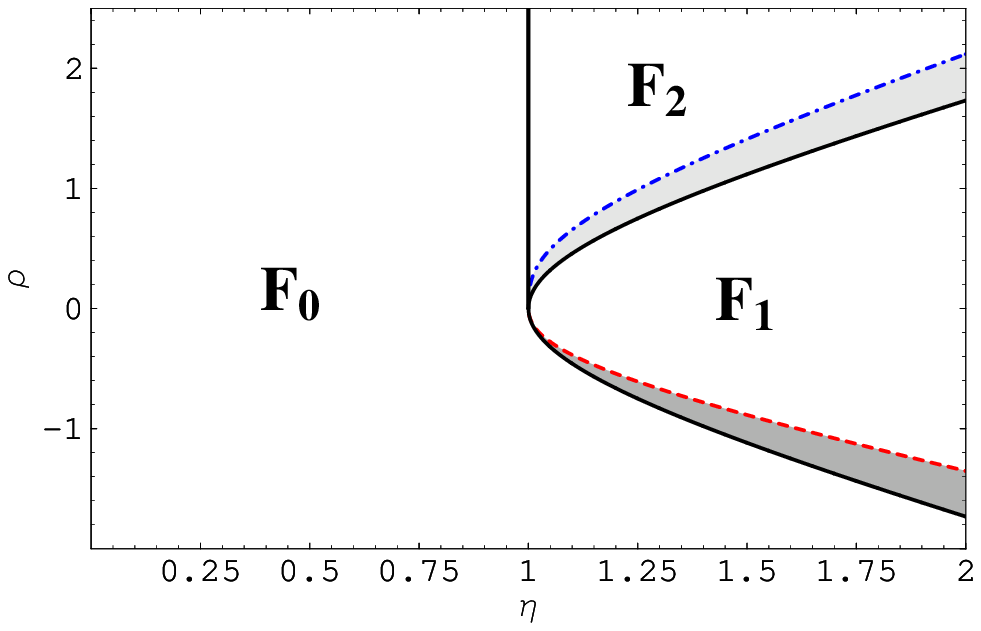}
\includegraphics[width=0.5\textwidth]{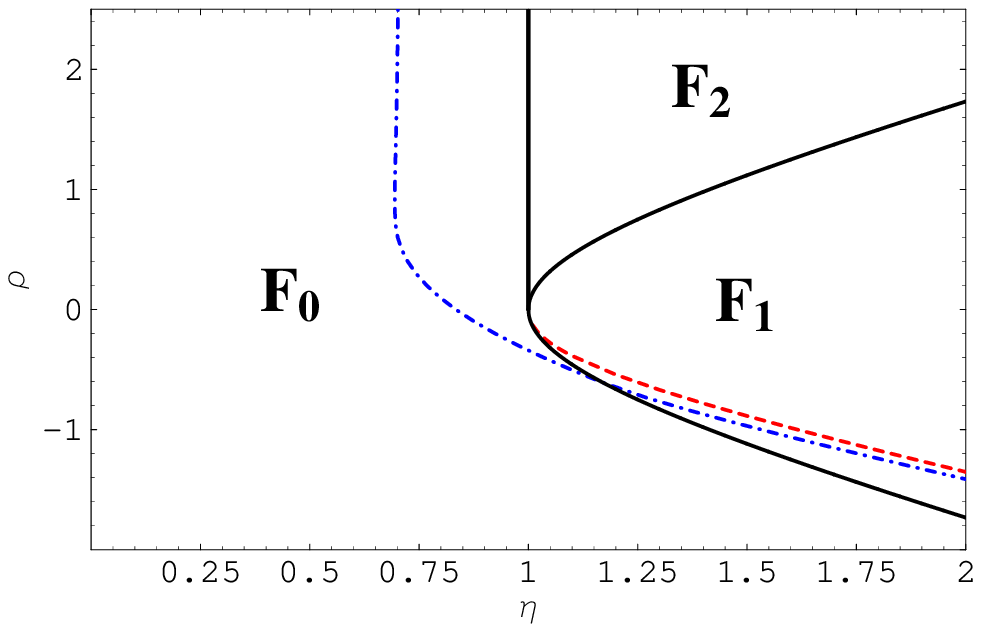}}
\caption{Different topologies of the effective Fermi surfaces, and stability conditions.}
\label{figglobal}
\end{center}
\end{figure}

We consider three cases distinguished by how many zeros quasiparticle dispersion has, $\varepsilon_k^-=\sqrt{(k^2/2m-\bar{\mu})^2+\Delta^2}-\delta\mu$, where $\mu$ is the average chemical potential and $\delta\mu$ is the difference between potentials, Fig.\ \ref{figoccupation}. In this case, number of zeros is equivalent to number of the Fermi surfaces. No zeros corresponds to the fully gapped state. In case of one Fermi surface, momenta outside the Fermi surface contribute to the pairing, while the excess of fermions resides inside the Fermi ball. In case of two Fermi surfaces, the excess of fermions resides between the two Fermi surfaces in momentum space. As was noticed by Son, apart from number of zeros, dispersion can have two different characteristic behaviors, distinguished by the position of the minimum. Minimum is located at nonzero momentum for positive $\bar{\mu}$, and corresponds to BCS, and minimum shifts to $p=0$ for negative $\mu$, and corresponds to BEC. This behavior manifests itself in stability conditions. 
We depict different topologies which are distinguished by the number of Fermi surfaces on the phase diagram in dimensionless average chemical potential $\bar{\mu}/\Delta$ and difference in chemical potentials $\delta\mu/\Delta$, the gap is the energy scale, in Fig.\ \ref{figglobal} regions between the solid lines.  We have fully gapped $F_0$, and gapless states with one $F_1$ and two $F_2$ Fermi surfaces. At small mismatch,  there is a fully gapped state $F_0$, which is at positive $\bar{\mu}$ and $\delta\mu<\Delta$, BCS, and at small and negative $\bar{\mu}$, $\bar{\mu}<-\sqrt{\delta\mu^2-\Delta^2}$, BEC. Increasing mismatch, when mismatch is at least larger than the gap, there is a gapless state with one Fermi surface $F_1$ when $\mu$ is around the Feshbach resonance, $-\sqrt{\delta\mu^2-\Delta^2}<\bar{\mu}<\sqrt{\delta\mu^2-\Delta^2}$, and gapless state with two Fermi surfaces $F_2$ for positive $\mu$ restricted from below, $\sqrt{\delta\mu^2-\Delta^2}<\bar{\mu}$. We therefore expect that $F_1$ exists at strong coupling, while $F_2$ probably exists only at weak coupling.

The question we are solving here is, "What is the ground state in a degenerate Fermi system with asymmetric number densities of fermions?". We avoid solving for the ground state explicitly. Instead, we check stability criteria, positive definite eigenvalues of screening masses and number susceptibility. If stability conditions are satisfied, homogeneous superconducting gapless state is indeed the ground state in this parameter range. If stability conditions are not satisfied, the alternative state may be realized as the ground state. These include LOFF, spatially separated mixtures, normal (not superconducting) states.

We analyze stability conditions, Eq.\ (\ref{screening}) and Eq.\ (\ref{chidef}), in all parameter space $(\bar{\mu},\delta\mu)$, and depict stable/unstable regions together with topology regions, Fig.\ \ref{figglobal} left panel. Debye mass e.v. are positive in all parameter space, hence Debye mass does not impose a constraint. All entries for the Meissner mass matrix are the same, it is trivial to diagonalize. We obtain\cite{1} $m_M^2=0$ corresponding to the unbroken sector $U(1)_{\alpha_1-\alpha_2}$, and $m_M^2=2L$ corresponding to the broken group $U(1)_{\alpha_1+\alpha_2}$, where
$L=\tilde{I}-\frac{\rho_+^{3/2}+\rho_-^{3/2}}{2\eta\sqrt{\eta^2-1}}$,
which defines the dashed-dotted (blue online) curve in Fig.\ \ref{figglobal} left panel. It renders all states between the dashed-dotted curve and the solid vertical line unstable. There is a strip left in gapless superconductor state with two Fermi surfaces $F_2$ which is stable, gray area. $F_0$ and $F_1$ are stable with respect to $m_M^2$ everywhere. In magnetic sector, mixing does not depend on chemical potentials and it is defined by the pattern of symmetry breaking. In electric sector, mixing depends on chemical potentials, e.g. in QCD mixing depends on $\delta\mu$. Mixing in electric and magnetic sectors is the same only for the fully gapped case.

Analyzing the number susceptibility matrix, Eq.\ (\ref{chidef}), we obtain\cite{1} the
expression defining the sign of e.v., which is very similar in structure to
that for the Meissner mass, see expression for $L$,
$R=I-\frac{\rho_+^{1/2}+\rho_-^{1/2}}{2\eta\sqrt{\eta^2-1}}$.
Here, we defined
$I\equiv I_{\rho}(0,\infty)-I_{\rho}(\sqrt{\rho_-},\sqrt{\rho_+})$ and
$\tilde{I}\equiv \tilde{I}_{\rho}(0,\infty)-
\tilde{I}_{\rho}(\sqrt{\rho_-},\sqrt{\rho_+ } )$ with elliptic
integrals
$I_{\rho}(a,b)\equiv  \int_a^b dx\,x^2/[(x^2-\rho)^2+1]^{3/2}$ and
$\tilde{I}_{\rho}(a,b)\equiv \int_a^b dx\,x^4/[(x^2-\rho)^2+1]^{3/2}$,
and $\rho_{\pm}=\rho\pm\sqrt{\eta^2-1}$ are zeros of $\varepsilon_k^{-}=0$ with
$\rho=\bar{\mu}/\Delta$, $\eta=\delta\mu/\Delta$. In expression for $R$
one should put $\rho_{-}=0$ when applied to $F_1$ region. It defines the dashed
(red online) line, and it renders all states between dashed and solid vertical
lines unstable. Thus all $F_2$ states are unstable,and there is a strip in $F_1$
which is stable, dark gray area. $F_0$ is stable everywhere. Stable states with
respect to $\chi$ correspond to local maximum of pressure.

We obtained stable regions which are local maxima of pressure, Fig.\ \ref{figglobal} left panel. Now we consider global maxima, Fig.\ \ref{figglobal} right panel. For this we compare pressure of the superconducting and normal states. Superconducting states which pressure is higher than that of the normal state are stable, $\Delta p=p_s-p_n>0$. The dashed-dotted (blue on line) line is $\Delta p=0$, it renders all states above and to the right of it unstable. All unstable regions with respect to the screening masses and number susceptibility are subset of unstable region with respect to the pressure. In a weak coupling, the vertical dashed-dotted line reproduces the known Clogston limit, $\delta\mu=\Delta/\sqrt{2}$, above which BCS is unstable. The global stability line cuts through the stable strip of $F_1$ state, below is a stable superconducting state, above is a metastable state. Both lines coincide at large mismatches.

Currently, experiments are being performed with unequal mixtures of fermions to map superfluid regions as a function of population imbalance, interaction strength and temperature. The experimental signature of superfluidity is the existence of vortices, which prove phase coherence in a sample.

\end{document}